\documentstyle[prd,aps,floats]{revtex}

\def\C{{\cal C}}
\def\P{{\cal P}}
\def\R{{\cal R}}
\def\S{{\cal S}}
\def\be{\begin{equation}}
\def\ee{\end{equation}}
\def\bea{\begin{eqnarray}}
\def\eea{\end{eqnarray}}

\begin{document}
\preprint{PU-ICG-02/06, astro-ph/0205253 v2} \draft

%
%
%
\renewcommand{\topfraction}{0.99}
\renewcommand{\bottomfraction}{0.99}
\twocolumn[\hsize\textwidth\columnwidth\hsize\csname
@twocolumnfalse\endcsname

\title{An Observational Test of Two-field Inflation}
\author{David Wands$^1$, Nicola Bartolo$^{2,3,4}$,
Sabino Matarrese$^{2,3}$ and Antonio Riotto$^3$}
\address{(1) Institute of Cosmology and Gravitation, University of
Portsmouth, Portsmouth PO1 2EG,~~~U.~K.}
\address{(2)  Dipartimento di Fisica di Padova ``G. Galilei'',
Via Marzolo 8, Padova I-35131, Italy}
\address{(3)  INFN, Sezione di Padova, Via Marzolo 8, Padova I-35131, Italy}
\address{(4) Astronomy Centre, University of Sussex, Brighton, BN1
  9QJ,~~~U.~K.}

\date{\today}
\maketitle
\begin{abstract}
  We study adiabatic and isocurvature perturbation spectra
produced by a period of cosmological inflation driven by two
scalar fields.
  We show that there exists a model-independent consistency
condition for all two-field models of slow-roll inflation, despite
allowing for model-dependent linear processing of curvature and
isocurvature perturbations during and after inflation on
super-horizon scales.
  The scale-dependence of all spectra are determined solely in
terms of slow-roll parameters during inflation and the dimensionless
cross-correlation between curvature and isocurvature perturbations. We
present additional model-dependent consistency relations that may be
derived in specific two-field models, such as the curvaton scenario.
\end{abstract}

\pacs{PACS numbers: 98.80.Cq \hfill Preprint PU-ICG/02-06,
astro-ph/0205253 v2}

\vskip2pc]

\section{Introduction}

The  primary interest in inflation \cite{LLbook,review}
is as a mechanism to explain the
origin of structure in the Universe from vacuum fluctuations in an
early inflationary era which are swept up to arbitrarily large
scales. The simplest inflationary models predict an almost
scale-invariant spectrum of Gaussian, adiabatic density
perturbations. Such a spectrum was already known as a likely model
of structure formation long before inflation was proposed.
Inflation also predicts a spectrum of gravitational waves or
`tensor' fluctuations. In single-field models of slow-roll
inflation there is a consistency condition between the slope of
the spectrum of tensor perturbations and the ratio of
tensor to scalar metric fluctuations \cite{LLbook}. The observational
confirmation of such a prediction is one of the holy grails of
modern cosmology.

Recent studies of multi-field models of inflation have
-- however -- threatened to destroy this appealing theoretical
prediction. Indeed, entropy perturbations generated in additional
light fields can alter the evolution of the curvature perturbation
even on large scales. This additional source for the late-time scalar
curvature perturbation breaks the single-field consistency relations,
yielding only an upper bound on the tensor-scalar
ratio~\cite{PolSta,SS,GBW}.

Consistency relations have recently been derived for adiabatic and
entropy perturbations during two-field inflation \cite{BMR}.
One of these relations explicitly shows how the single-field
consistency condition is modified through the cross-correlation
between the entropy and the curvature perturbations. On the other hand,
the model dependent nature of reheating at the end of inflation
makes it impossible to quantify the late-time amplitude of entropy
perturbations solely in terms of the evolution during inflation.

In this paper we introduce a model-independent description of the
coupling between adiabatic and entropy perturbations both during and
after inflation in order to relate late-time observables to
perturbation spectra generated during inflation. We show that the
scale dependence of all spectra depend only on quantities at
horizon-crossing during slow-roll inflation and the cross-correlation
between curvature and isocurvature perturbations.  This
cross-correlation generically arises in inflation models for the
origin of curvature and isocurvature
perturbations~\cite{Langlois,Gordonetal,BMR}. If it can be determined
observationally~\cite{LR,BMTII,Durrer,AGWS} then one can reconstruct
the scalar curvature perturbation at horizon-crossing. We are thus
able to derive a generalised consistency condition for the observed
tensor-scalar ratio that holds in {\em all} two-field models of
slow-roll inflation.

The paper is organized as follows. In section II we introduce the
notion of transfer functions and deal with slow-roll inflation in
section III. In sections IV and V we compute the power spectra at
horizon crossing and at late times, respectively, and derive the
consistency relation holding for all two-field models of inflation. In
section VI we present additional relations that hold for restricted
classes of slow-roll models. Finally, in section VII we draw our
conclusions.

\section{Transfer functions}

We will first construct dimensionless quantities to describe the
instantaneous adiabatic (curvature) and entropy (isocurvature)
perturbations both during and after inflation. We can then relate
their values on large scales during and after inflation by a
model-dependent transfer matrix, whose general form will be given in
Eq.~(\ref{defT}).

During the conventional radiation-dominated era (after inflation)
there is a conventional gauge-invariant definition for the
large-scale density/curvature
perturbation~\cite{BST,Bardeen88,Wandsetal}
\begin{equation}
\label{defR}
 \R = \psi + {H\delta\rho\over \dot\rho} \,,
\end{equation}
where $\psi$ is the gauge-dependent curvature perturbation and
$\delta\rho$ the total density perturbation in that gauge. The
isocurvature/entropy perturbation is usually defined as the perturbation
in the ratio of the matter and photon number densities
\begin{equation}
\label{defSBg}
 \S = {\delta n_m\over n_m} - {\delta n_\gamma\over
n_\gamma} =
 -3\left( {H\delta\rho_ m \over \dot\rho_ m} - {H\delta\rho_\gamma \over
 \dot\rho_\gamma} \right)
  \, ,
\end{equation}
which is naturally gauge-independent.

The `primordial' adiabatic and isocurvature perturbations on
cosmological scales ($1-10^4$Mpc) are usually defined in terms of
the early-time/large-scale limit deep in the radiation dominated
era~\cite{isocurv}, e.g., around the epoch of primordial
nucleosynthesis ($T\sim10^{10}$K). The power spectrum and
cross-correlation of the primordial adiabatic and isocurvature
perturbations on cosmological scales can then be constrained by
observations, such as the cosmic microwave background angular power
spectrum\cite{CMB,LR,BMTII,Durrer,AGWS}.

During two-field inflation the general definition of the curvature
perturbation (\ref{defR}) yields
\begin{equation}
 \label{defRi}
\R \simeq {H \left(
\dot\phi\delta\phi+\dot\chi\delta\chi\right)\over
 \dot\phi^2+\dot\chi^2}
 \,.
\end{equation}
where $\simeq$ denotes equality in the slow-roll
approximation\footnote{Although the curvature and field
perturbations are, in general, gauge-dependent,  this
gauge-dependence can for most purposes be neglected at leading
order in the slow-roll approximation.
For definiteness one can take all field perturbations to be
evaluated in the spatially-flat gauge~\cite{KS}.}.
The generalised entropy (isocurvature) perturbation is given
by~\cite{GBW,Gordonetal}
\begin{equation}
\label{defSi}
 \S = {H \left(
\dot\phi\delta\chi-\dot\chi\delta\phi\right)\over
 \dot\phi^2+\dot\chi^2}
   \,.
\end{equation}
%
%
As $\S$ is not directly observable during inflation, its
normalisation is somewhat arbitrary. This particular choice keeps
the subsequent analysis of power spectra simpler by giving
curvature and isocurvature spectra equal power at horizon-crossing
[see. Eq.~(\ref{PR*})]. A different choice for the normalisation
of $\S$ would lead to a different overall factor multiplying the
transfer functions $T_{\R\S}$ and $T_{\S\S}$ in Eq.~(\ref{defT}).

In order to relate the initial curvature and entropy perturbations
(\ref{defRi}) and (\ref{defSi}) generated by a period of inflation
in the very early universe, to the observable curvature and
entropy perturbations (\ref{defR}) and (\ref{defSBg}) at much
later cosmic times, we need to model the evolution on large
(``super-horizon'') scales.
We will work in a large-scale limit where the divergence of the
velocity field and shear can be neglected so that the local
dynamics are those of a homogeneous and isotropic FRW
model~\cite{Wandsetal}. During slow-roll inflation this becomes a
good approximation soon after a mode leaves the Hubble-scale
($k<aH$), and it remains valid up until the mode re-enters the
Hubble-scale during the subsequent radiation or matter dominated
eras. Adiabatic perturbations correspond to perturbations which
locally follow the same trajectory in phase-space as the
unperturbed background, whereas entropy perturbations correspond
to perturbations off the background trajectory~\cite{Wandsetal}.

The curvature perturbation $\R$ remains constant for purely adiabatic
perturbations in the large-scale limit simply as a consequence of
local energy conservation~\cite{Wandsetal}, irrespective of the
physical processes going on during inflation or reheating. Purely
adiabatic perturbations can never generate entropy perturbations on
large scales, but entropy perturbations (specifically a non-adiabatic
pressure perturbation or energy transfer) can change the curvature
perturbation. Moreover, the entropy perturbation itself can evolve on
large scales for imperfect fluids\footnote{For the special case of two
  non-interacting perfect fluids, such as matter and radiation, the
  isocurvature perturbation defined in Eq.~(\ref{defSBg}) is constant
  on large scales.}.
One can thus argue on very general grounds~\cite{Wandsetal} that
the time dependence of adiabatic and entropy perturbations in the
large-scale limit can always be described by
\begin{eqnarray}
\label{dR}
\dot{\R} = \alpha H \S \,, \,\,\,\,
\quad
\label{dS}
\dot{\S} = \beta H \S \,,
\end{eqnarray}
where $\alpha$ and $\beta$ are in general time-dependent
dimensionless functions.
The explicit form of the interaction between the curvature and
entropy perturbations has recently been explicitly demonstrated in
the case of interacting scalar
fields~\cite{Gordonetal,Hwangfields,Nibbelink} and non-interacting
fluids~\cite{Hwangfluids}.

Integrating Eqs.~(\ref{dR})  over time we can obtain the general
form of the transfer matrix relating curvature and entropy
perturbations generated when a given mode is stretched outside the
Hubble scale during inflation ($k=aH$, denoted by an asterisk) to
curvature and entropy perturbations at some later time~\cite{AGWS}:
\begin{equation}
\label{defT}
\left(
\begin{array}{c}
{\R} \\ {\S}
\end{array}
\right) = \left(
\begin{array}{cc}
1 & {T}_{\R\S} \\ 0 & {T}_{\S\S}
\end{array}
\right) \left(
\begin{array}{c}
\R \\\S
\end{array}
\right)_* \,,
\end{equation}
where
\begin{eqnarray}
 \label{TRS}
T_{\R\S}(t_*,t) &=& \int^t_{t_*} \alpha(t') T_{\S\S}(t_*,t') H(t')
dt' \,,
\nonumber\\
 \label{TSS}
T_{\S\S}(t_*,t) &=& \exp \left( \int^t_{t_*} \beta(t') H(t')
dt' \right) \,.
\end{eqnarray}

Although the evolution in the large-scale limit is independent of
scale (by definition), the transfer functions $T_{\R\S}$ and
$T_{\S\S}$ are implicit functions of scale due to their dependence
upon $t_*(k)$.
The scale-dependence of the transfer functions is proportional to
\begin{eqnarray}
  \label{dTRS}
 H_*^{-1} {\partial T_{\R\S}\over \partial t_*}
 &=& -\alpha_* -\beta_* T_{\R\S}
 \,, \nonumber \\
  \label{dTSS}
 H_*^{-1} {\partial T_{\S\S} \over \partial t_*} &=&
-\beta_* T_{\S\S} \,.
\end{eqnarray}
Thus the scale-dependence of the transfer functions is determined
by $\alpha_*$ and $\beta_*$ which describe the evolution of the
curvature and entropy fluctuations at Hubble-exit during
inflation. We shall now show how these can be related to the
dimensionless slow-roll parameters during inflation.

\section{Slow-roll inflation}

We will consider inflation driven by
minimally-coupled real
scalar fields $\phi$ and $\chi$ with arbitrary potential
$V(\phi,\chi)$.
Homogeneous scalar fields in a flat FRW cosmology obey the coupled
Klein-Gordon equations
\begin{eqnarray}
\ddot\phi + 3H\dot\phi = - V_{\phi} \,,\nonumber\\
\ddot\chi + 3H\dot\chi = - V_{\chi} \,,
\end{eqnarray}
where $V_x$ denotes $\partial V/\partial x$, subject to the
Friedmann constraint
\begin{equation}
H^2 = {8\pi G\over 3} \left( {1\over2} \dot\phi^2 + {1\over2}
\dot\chi^2 + V \right) \,.
\end{equation}
Fields interact both through their explicit interaction potential
and gravitationally.

Inhomogeneous but linear perturbations about the non-linear but
homogeneous background solutions obey the perturbed Klein-Gordon
equations which can be written as~\cite{TN}
\begin{eqnarray}
\ddot{\delta\phi} + 3H \dot{\delta\phi}
 + \left[ {k^2\over a^2} + V_{\phi\phi} - {8\pi G\over a^3}
 {d\over dt} \left( {a^3\over H} \dot\phi^2 \right)
 \right] \delta\phi \nonumber\\
  =
 -\left[ V_{\phi\chi} - {8\pi G\over a^3}
 {d\over dt} \left( {a^3\over H} \dot\phi\dot\chi
 \right)  \right]  \delta\chi \,,\nonumber\\
\ddot{\delta\chi} + 3H \dot{\delta\chi}
 + \left[ {k^2\over a^2} + V_{\chi\chi} - {8\pi G\over a^3}
 {d\over dt} \left( {a^3\over H} \dot\chi^2 \right)
 \right] \delta\chi \nonumber\\
 = -\left[ V_{\chi\phi}  - {8\pi G\over a^3}
 {d\over dt} \left( {a^3\over H} \dot\chi\dot\phi
 \right)  \right] \delta\phi\,.
\end{eqnarray}
where the field perturbations are defined in the spatially-flat
gauge~\cite{KS,Sasaki,Hwang}.

In the standard approach~\cite{SS,GBW,BMR} one defines five
slow-roll parameters, two describing the slope of the potential
\begin{equation}
\epsilon_\phi \equiv {1\over 16\pi G} \left( {V_\phi\over V}
 \right)^2 \, , \quad
\epsilon_\chi \equiv {1\over 16\pi G} \left(
 {V_\chi\over V} \right)^2 \, ,
\end{equation}
and three describing the curvature
\begin{eqnarray}
\eta_{\phi\phi} \equiv {1\over 8\pi G} \left( {V_{\phi\phi}\over
 V} \right) \, , \quad
\eta_{\chi\chi} \equiv {1\over 8\pi G} \left(
 {V_{\chi\chi}\over V} \right) \, , \nonumber\\
\eta_{\phi\chi} \equiv {1\over 8\pi G} \left( {V_{\phi\chi}\over
 V} \right) \, . \qquad
\end{eqnarray}
The slow-roll equations give an approximate solution for the
growing mode solution when
max$\{\epsilon_i,|\eta_{ij}|\}\ll 1$.
The slow-roll solutions for the homogeneous background are given
by
\begin{equation}
\dot\phi^2 \simeq {2\over3} \epsilon_\phi V \,,
\quad
\dot\chi^2 \simeq {2\over3} \epsilon_\chi V \,.
\end{equation}
where $\simeq$ denotes equality in the slow-roll
approximation while linear perturbations on large-scales ($k\ll
aH$) obey
\begin{eqnarray}
H^{-1} \dot{\delta\phi} &\simeq&
 \left( 2\epsilon_\phi -\eta_{\phi\phi} \right) \delta\phi
 + \left( \pm 2\sqrt{\epsilon_\phi\epsilon_\chi} - \eta_{\phi\chi} \right) \delta\chi \,,\\
\nonumber
 H^{-1} \dot{\delta\chi} &\simeq&
 \left( 2\epsilon_\chi -\eta_{\chi\chi} \right) \delta\chi
 + \left( \pm 2\sqrt{\epsilon_\chi\epsilon_\phi} - \eta_{\phi\chi} \right) \delta\phi
 \,.
\end{eqnarray}
The $\eta_{ij}$ slow-roll parameters represent the explicit
interaction via the potential $V(\phi,\chi)$, while $\epsilon_i$
terms are due to the gravitational coupling.

We will adopt the approach of Gordon {\it et al.} \cite{Gordonetal}
and perform a local field rotation to identify the instantaneous
adiabatic and entropy perturbations along and orthogonal to the
background trajectory in field-space:
\begin{eqnarray}
\label{rotate}
\delta\sigma = \cos\theta \delta\phi + \sin\theta \delta\chi
 \,,\nonumber \\
\delta s = -\sin\theta \delta\phi + \cos\theta \delta\chi
 \, ,
\end{eqnarray}
where
$\tan\theta=\dot\chi/\dot\phi\simeq\pm\sqrt{\epsilon_\chi/\epsilon_\phi}$.
This approach can be readily extended to include multiple scalar
fields and non-minimal coupling~\cite{Nibbelink}.
The curvature and entropy perturbations, defined in
Eqs.~(\ref{defRi}) and~(\ref{defSi}) during inflation, then take
the simple form
\begin{equation}
\label{defRSinf}
 \R \simeq {H\delta\sigma \over \dot\sigma} \,,
\quad
 \S = {H\delta s \over \dot\sigma} \,
\end{equation}

The local field rotation (\ref{rotate}) allows one of the
slow-roll parameters to be eliminated, in this case the slope
orthogonal to the trajectory, $\epsilon_s\simeq0$ in slow-roll,
and we are left with four parameters: one describing the slope of
the potential
\begin{equation}
\epsilon \equiv {1\over 16\pi G} \left( {V_\sigma\over V}
 \right)^2 \simeq \epsilon_\phi + \epsilon_\chi \,
 ,
\end{equation}
and three describing the curvature
\begin{eqnarray}
\eta_{\sigma\sigma}
 &=& \eta_{\phi\phi} \cos^2\theta
  + 2 \eta_{\phi\chi}\cos\theta\sin\theta
   + \eta_{\chi\chi} \sin^2\theta \, ,
 \nonumber\\
\eta_{\sigma s}
 &=& ( \eta_{\chi\chi} - \eta_{\phi\phi} ) \sin\theta\cos\theta
  + \eta_{\phi\chi} ( \cos^2\theta - \sin^2\theta ) \,,
  \\
\eta_{ss}
  &=& \eta_{\phi\phi} \sin^2\theta
   - 2 \eta_{\phi\chi} \sin\theta\cos\theta
    + \eta_{\chi\chi} \cos^2\theta \, . \quad
 \nonumber
\end{eqnarray}
Alternatively we could choose to diagonalise the mass matrix,
$V_{ij}$, to eliminate one $\eta$-term and have two slopes and two
curvature parameters. Either way we see that the local evolution
of the fields and their perturbations at any instant can be
described by {\em four} slow-roll parameters.

The background slow-roll solution is then given by
\begin{equation}
\dot\sigma^2 \simeq {2\over3} \epsilon V \,,
\quad
H^{-1}\dot\theta \simeq -\eta_{\sigma s} \, ,
\end{equation}
while the perturbations obey

\begin{eqnarray}
H^{-1} \dot{\delta\sigma} &\simeq&
 \left( 2\epsilon -\eta_{\sigma\sigma} \right) \delta\sigma
 - 2\eta_{\sigma s}\delta s \,,
  \nonumber\\
H^{-1} \dot{\delta s} &\simeq& -\eta_{ss}\delta s \,.
\end{eqnarray}
The entropy field perturbation $\delta s$ evolves independently of
the adiabatic field perturbation $\delta\sigma$ on large scales.
However the large-scale entropy perturbations do affect the
evolution of the adiabatic perturbations when $\eta_{\sigma s}\neq
0$.

In terms of the dimensionless perturbations $\R$ and $\S$ we have
\begin{eqnarray}
\label{dotR}
 \dot\R &\simeq& - 2\eta_{\sigma s} H \S \,,
 \nonumber\\
\label{dotS}
 \dot\S &\simeq& \left( -2\epsilon
+\eta_{\sigma\sigma} -
 \eta_{ss} \right) H \S \,.
\end{eqnarray}
which provides a specific example of the more general form for the
evolution of curvature and entropy perturbations given in
Eqs.~(\ref{dR}).
In particular the scale-dependence of the integrated transfer
functions, Eqs.~(\ref{dTRS}), can be written in terms of the
slow-roll parameters when the mode crosses outside the
Hubble-scale:
\begin{eqnarray}
\label{sralpha} \alpha_* &\simeq& - 2\eta_{\sigma s}
 \,,\nonumber\\
\label{srbeta} \beta_* &\simeq& -2\epsilon +\eta_{\sigma\sigma} -
 \eta_{ss} \,.
\end{eqnarray}


\section{Initial power spectra}

Weakly-interacting, light fields acquire a spectrum of vacuum
fluctuations at Hubble-crossing ($k=a_*H_*$) \cite{LLbook}
\begin{equation}
{\cal P}_{\delta\phi}|_* \simeq {\cal P}_{\delta\chi}|_* \simeq
\left( {H_* \over 2\pi} \right)^2 \,,
\end{equation}
which describe independent Gaussian random fields, {\it i.e.} zero
cross-correlation
\begin{equation}
{\C}_{\delta\phi,\delta\chi}|_* = 0 \,.
\end{equation}

The local rotation (\ref{rotate}) to the instantaneous adiabatic and
entropy field perturbations, gives
\begin{eqnarray}
{\cal P}_{\delta\sigma}|_* \simeq
 {\cal P}_{\delta s}|_* \simeq
  \left( {H_* \over 2\pi} \right)^2 \,,\nonumber\\
{\C}_{\delta\sigma,\delta s}|_* = 0 \,. \quad
\end{eqnarray}

Hence, using Eq.~(\ref{defRSinf}), the adiabatic and entropy power
spectra at Hubble-crossing are given by
\begin{equation}
\label{PR*} \P_\R|_* \simeq \P_\S|_* \simeq \left( {H^2 \over
2\pi\dot\sigma} \right)^2_* \simeq {8\over3\epsilon} {V_*\over
M_P^4} \,.
\end{equation}
Although,
 as explained earlier,
the normalisation of the dimensionless entropy perturbation during
inflation is arbitrary, it proves convenient to use that given in
Eq.~(\ref{defSi}) so that $\R$ and $\S$ have equal power at
Hubble-crossing.

The spectral tilts (defined by $n_x\equiv {d\ln \P_x / d \ln
k}$) are given by
\begin{equation}
\label{nR*}
n_\R|_* \simeq
 n_\S|_*  \simeq - 6\epsilon + 2\eta_{\sigma\sigma}\,.
\end{equation}
%

Gravitational waves are generated with a spectrum~\cite{LLbook}
\begin{equation}
\label{PT*}
\P_T|_* \simeq {128\over 3} {V_* \over M_P^4} \,
\end{equation}
and spectral tilt
\begin{equation}
\label{nT*} n_T|_* \simeq -2\epsilon \,.
\end{equation}

A key observation is that the tensor-scalar ratio at
Hubble-crossing, even in multi-field slow-roll inflation, can be
given from Eqs.~(\ref{PR*}), (\ref{PT*}) and~(\ref{nT*}) as
\begin{equation}
\label{TR*} \left( {\P_T\over \P_\R} \right)_* \simeq 16 \epsilon
\simeq -8 n_T|_* \,.
\end{equation}

\section{Final power spectra}

Applying the transfer matrix (\ref{defT}) to the initial scalar spectra we
obtain the resulting curvature and isocurvature power spectra at the
start of the conventional radiation-dominated era:
\begin{eqnarray}
\label{PR}
\P_{\R} &=& \left( 1 + T_{\R\S}^2 \right) \P_\R|_* \,,\\
\label{PS}
\P_{\S} &=& T_{\S\S}^2 \P_\R|_* \,,\\
\label{CRS}
 \C_{\R\S} &=& T_{\R\S} T_{\S\S} \P_\R|_* \,.
\end{eqnarray}
A dimensionless measure of the correlation can be defined in terms of
a correlation angle $\Delta$ such that
\begin{equation}
\label{defDelta} \cos\Delta \equiv {\C_{\R\S} \over \P_{\R}^{1/2}
  \P_{\S}^{1/2}} \simeq {T_{\R\S} \over \sqrt{1+T_{\R\S}^2}} \,.
\end{equation}
Note that the scalar metric perturbation at Hubble-crossing can
thus be reconstructed from the observed curvature perturbation at
late times and the cross-correlation angle:
\begin{equation}
\label{reconstruct}
\P_{\R}|_* \simeq \P_{\R} \sin^2\Delta \,.
\end{equation}

The tensor perturbations, in contrast to the scalar perturbations,
remain `frozen-in' on large scales, and decoupled from the scalar
perturbations at linear order. Thus the primordial perturbation
spectrum for gravitational waves is given by Eqs.~(\ref{PT*})
and~(\ref{nT*})
\begin{equation}
\label{PT} \P_T = \P_T|_* \,,\quad n_T = n_T|_* \,.
\end{equation}

The consistency condition for the tensor-scalar amplitudes at
Hubble-crossing can thus be rewritten using Eqs.~(\ref{TR*}),
(\ref{reconstruct}) and~(\ref{PT}) as a consistency relation
between the tensor-scalar amplitudes
at late times:
%
\begin{equation}
\label{consistency} {\P_T\over \P_\R} \simeq - 8 n_T \sin^2\Delta
\,.
\end{equation}
Any two-field model of slow-roll inflation predicts that this
relation should hold between quantities which are directly
observable, so long as the amplitude of the isocurvature and
tensor perturbations prove to be large enough.
This relation was first obtained in Ref. \cite{BMR} just \emph{at the
end} of an inflationary period where two scalar fields are present
(see Eq. (59) of Ref. \cite{BMR}). We have shown that this
consistency relation (\ref{consistency}) between the tensor-scalar
amplitudes is not modified by any linear processes the matter
perturbations may undergo between the end of inflation and some
later time $t$.

The scale-dependence of the final scalar power spectra depends
both on the scale dependence of the initial spectra ($n_\R|_*$)
and on the transfer functions $T_{\R\S}$ and $T_{\S\S}$.
The spectral tilts are given from Eqs.~(\ref{PR}--\ref{CRS})
by\footnote{In this notation, the spectral index for adiabatic
scalar perturbations is conventionally given as $n=1+n_\R$ so that
a scale-invariant (Harrison-Zel'dovich) spectrum corresponds to
$n_\R=0$.}
\begin{eqnarray}
 \label{gentilt}
n_\R &=& n_\R|_* + H_*^{-1} (\partial{T}_{\R\S}/\partial t_*) \sin 2\Delta \,,
 \nonumber\\
n_\S &=& n_\R|_* + 2 H_*^{-1} (\partial\ln{T}_{\S\S}/\partial t_*) \,,\\
n_\C &=& n_\R|_* + H_*^{-1} \left[ (\partial{T}_{\R\S}/\partial
t_*) \tan\Delta + (\partial\ln{T}_{\S\S}/\partial t_*) \right] \,,
 \nonumber
\end{eqnarray}
where we have used Eq.~(\ref{defDelta}) to eliminate $T_{\R\S}$ in
favour of the observable correlation angle $\Delta$.
Substituting in Eq.~(\ref{nR*}) for the tilt at Hubble-exit, and
Eqs.~(\ref{dTRS}) and (\ref{srbeta}) for the scale-dependence of
the transfer functions, we obtain
\begin{eqnarray}
\label{srtilts}
n_{\R} &\simeq& -(6-4\cos^2\Delta) \epsilon \nonumber\\
&&  + 2\left( \eta_{\sigma\sigma}\sin^2\Delta + 2\eta_{\sigma
 s}\sin\Delta\cos\Delta + \eta_{ss}\cos^2\Delta \right)
\,,\nonumber\\
n_\S &\simeq& -2\epsilon + 2\eta_{ss} \,,\\
n_\C &\simeq& -2\epsilon + 2\eta_{ss} + 2\eta_{\sigma
  s}\tan\Delta
 \nonumber\,.
\end{eqnarray}
We emphasize that although the overall amplitude of the transfer
functions $T_{\R\S}$ and $T_{\S\S}$ are dependent upon the
evolution after Hubble-exit, through reheating and into the
radiation dominated era, the spectral tilts of the resulting
perturbation spectra can be expressed solely in terms of the
slow-roll parameters at Hubble-crossing during inflation and the
correlation angle $\Delta$ which can (in principle) be determined
by observations.

\section{Model-dependent relations}

In addition to the consistency condition~(\ref{consistency}) that
applies to any slow-roll model of two-field inflation, there are
additional model-dependent relations that will hold for restricted
classes of two-field inflation.

In Ref.~\cite{BMR} a second consistency relation was derived for
curvature and entropy perturbations at the end of inflation using
the integrated slow-roll solutions:
\begin{equation}
n_\C-n_\S + {n_\R + n_\S - 2n_\C \over 2\sin^2\Delta} \approx 0
 \,.
\end{equation}
{}From Eqs.~(\ref{gentilt}) one sees that in fact this holds for all
models for which $H_*^{-1}\partial \ln T_{\S\S}/\partial t_*\approx0$.
This requires $\beta_*\approx0$ in Eq.~(\ref{dTSS}) which can be given
as a constraint on the slow-roll parameters at Hubble-exit by
Eq.~(\ref{srbeta}).

Another class of two-field inflation models are those in which the
curvature and isocurvature perturbations are effectively decoupled
around the time of Hubble-exit, $\alpha_*\approx0$ in
Eq.~(\ref{dR}).
In terms of slow-roll parameters at Hubble-exit this requires,
from Eq.~(\ref{sralpha}), that $\eta_{\sigma s}\approx0$. This
includes models in which only one scalar field evolves during
inflation, but where both fields play a significant dynamical role
during reheating or afterwards. {}From Eqs.~(\ref{srtilts}) we see
immediately that we have the constraint
\begin{equation}
 n_\S \approx n_\C \,.
\end{equation}

The curvaton model for the origin of cosmological structure
proposed in Ref.~\cite{curvaton,Moroi} falls into this class of models.
In the curvaton scenario the initial curvature perturbation at
Hubble-exit (or later) during inflation is taken to be negligible,
i.e., $\R_*\ll\R$. The curvature perturbation observed during the
subsequent radiation or matter dominated eras is supposed to be
due entirely to an initial isocurvature perturbation at
Hubble-exit. As can be seen from Eq.~(\ref{reconstruct}) this
implies that the resulting curvature and isocurvature
perturbations must be 100\% correlated, $\sin\Delta\approx0$.
Unfortunately Eq.~(\ref{consistency}) then shows that the
amplitude of tensor perturbations must be negligible,
$\P_T\approx0$. Instead we have a constraint solely in terms of
the scalar spectra from Eqs.~(\ref{srtilts}):
\begin{equation}
n_\R \approx n_\S \approx n_\C
 \,.
\end{equation}


\section{Conclusions}

In this paper we have shown how curvature and entropy perturbations
produced by any slow-roll model of two-field inflation can be related
to observable curvature and matter-isocurvature perturbation
spectra at late times.

The resulting amplitude and tilt of the spectra of curvature and
entropy perturbations and their correlation can be described by six
parameters which may in principle be determined observationally.
These six observables are determined by seven model parameters: the
Hubble rate during inflation, four dimensionless slow-roll parameters,
and two transfer functions $T_{\R\S}$ and $T_{\S\S}$ which are
dependent upon the detailed physics of reheating.
This situation is analogous to the case of single-field inflation
where the two observables (amplitude and tilt) of the adiabatic
curvature perturbation spectrum is determined by three model parameters:
the Hubble rate during inflation and two slow-roll parameters.

To break the degeneracy we require an observable spectrum of
gravitational waves produced during inflation, whose amplitude and
tilt gives two more observables, and hence the observationally
testable consistency relation~(\ref{consistency}), which is a
generalisation of the single-field relation~\cite{LLbook}.

Although the amplitude of the isocurvature and cross-correlation
spectra are dependent upon two transfer functions $T_{\R\S}$ and
$T_{\S\S}$ which are, a priori, unknowns, the correlation angle,
$\Delta$, is a direct measure of one of these, $T_{\R\S}$. This
enables one to quantify the contribution of non-adiabatic
perturbations to the late-time curvature and hence reconstruct the
original curvature perturbation spectrum at Hubble-exit from that
observed at late times.  A measure of the amplitude of the late-time
isocurvature amplitude then allows one to determine $T_{\S\S}$.

We have shown that the spectral tilts of the tensor and scalar spectra
can be written in terms of the four slow-roll parameters describing
the evolution at the time of Hubble-exit during inflation. This yields
additional consistency relations in specific models such as the
curvaton scenario~\cite{curvaton,Moroi}.

Finally we note that the relation (\ref{reconstruct}) between the
isocurvature correlation angle $\Delta$
and the change in the large-scale curvature assumes that only one
entropy mode exists at horizon-exit during inflation. In inflation
models with more than two light fields during inflation, an
additional uncorrelated entropy mode at horizon-crossing could
contribute to the isocurvature without affecting the curvature at
late times. Thus our generalised consistency relation
(\ref{consistency}) only applies to two-field models of inflation.
For three or more light fields during inflation we again have an
inequality
\begin{equation}
\label{inconsistency}
{\P_T\over \P_\R} \leq - 8 n_T \sin^2\Delta \,.
\end{equation}
At the same time, we have only considered one observable isocurvature
mode in the radiation-dominated era. The general cosmological
perturbation can include as many isocurvature modes as there are
distinguishable matter components~\cite{BMT}.
The correlation of these additional modes with the curvature could
enable one to reconstruct the curvature perturbation at horizon-exit
even in the presence of additional light fields during inflation.
In general one would expect to be able to find a tensor-scalar
consistency condition when there are as many observable perturbation
modes in the radiation dominated universe after inflation as there are
light fields during inflation.

\acknowledgments

NB was partially supported by a Marie Curie Fellowship of the European
Community programme HUMAN POTENTIAL under contract HPMT-CT-2000-00096.
DW is supported by the Royal Society.


\end{document}